\documentclass[aps,pra,reprint, amsmath, amssymb,superscriptaddress]{revtex4-1}

\usepackage{bm}
\usepackage[retainorgcmds]{IEEEtrantools}
\usepackage{graphicx}
\usepackage{mathrsfs}
\usepackage{amsmath}
\usepackage{amssymb}
\usepackage{color}
\usepackage{amsfonts}
\usepackage{nicefrac}

\newcommand{\ud}{\,\mathrm{d}}

\DeclareMathOperator{\tr}{tr}

\usepackage{pgfplots}

\definecolor{myorange}{RGB}{224,148,57}
\definecolor{myblue}{RGB}{68,151,151}
\definecolor{mypurple}{RGB}{158,67,149}
\definecolor{myblue2}{RGB}{205,205,222}
\definecolor{mynavy}{RGB}{48,50,120}

\tikzstyle{dashed1}= [dash pattern=on 6pt off 1pt]
\tikzstyle{dashed2}= [dash pattern=on 3pt off 1pt]
\tikzstyle{dashed3}= [dash pattern=on 3pt off 1pt on 1pt off 1pt]

\pgfplotsset{
scale only axis, 
mydef/.style={
legend cell align=left,
tick label style={font=\scriptsize},
label style={font=\scriptsize},
legend style={
fill=none,draw=none,
font=\scriptsize}
}}

\begin{document}

\title{Irreversibility at zero temperature from the perspective of the environemnt}
\date{\today}
\author{Jader P. Santos}
\affiliation{Instituto de F\'isica da Universidade de S\~ao Paulo,  05314-970 S\~ao Paulo, Brazil}
\author{Alberto L. de Paula Jr.}
\author{Raphael Drumond}
\affiliation{Departamento de F\'isica, Instituto de Ci\^encias Exatas, Universidade Federal de Minas Gerais, 30123-970, Belo Horizonte, Minas Gerais, Brazil}
\author{Gabriel T. Landi}
\email{gtlandi@if.usp.br}
\affiliation{Instituto de F\'isica da Universidade de S\~ao Paulo,  05314-970 S\~ao Paulo, Brazil}
\author{Mauro Paternostro}
\affiliation{Centre for Theoretical Atomic, Molecular and Optical Physics,
School of Mathematics and Physics, Queen's University Belfast, Belfast BT7 1NN, United Kingdom}

\begin{abstract}
We address the emergence of entropy production in the non-equilibrium process of an open quantum system from the viewpoint of the environment. By making use of a dilation-based approach akin to Stinespring theorem, we derive an expression for the entropy production that comprises two fundamental contributions. The first is linked to the rate of creation of correlations between system and environment whereas the second highlights the possibility for the environment to modify its state in light of its coupling to the system. 
Both terms are shown to be associated with irreversible currents within the system and the environment, which pinpoint the emergence of irreversibility in the Markovian limit. 
Finally, we discuss how such a change of perspective in the study of entropy production has fecund implications for the study of non-Markovian open-system dynamics. 
\end{abstract}
\maketitle{}

%
%
%
%

\emph{Introduction - }
Irreversibility is an emergent concept, stemming from the unavoidable loss of information that occurs during the evolution of systems containing a macroscopically large number of degrees of freedom. 
Despite considerable research for more than a century, several open questions remain as to the physical origins and implications of this concept.
Recently, however, this field has experienced a boom of advances, motivated partially by improvements in the experimental control of mesoscopic systems  \cite{Liphardt2002,Speck2007,Collin2005,Ciliberto2013,Roßnagel2015} and quantum technology platforms \cite{Myatt2000,Groeblacher2013}. 
This has allowed us, for the first time, to have direct control over the number of degrees of freedom \cite{Martinez2015} and the magnitude of classical and quantum fluctuations \cite{Batalhao2014,*Batalhao2015}, opening the way to directly experiment with the emergence of irreversibility. Such advances have also been accompanied by fundamental progress in our theoretical understanding of out-of-equilibrium thermodynamics, 
such as the discovery of fluctuation theorems \cite{Crooks1998,*Crooks2000,Jarzynski1997,*Jarzynski1997a,Jarzynski2004a,Tasaki2000,Seifert2012,Ford2012,Manzano2017a,Dorner2012}, 
 the role of quantum correlations \cite{Leandro2009,Popescu2006} and the interplay between thermodynamics and information \cite{Parrondo2015,Cottet2017,Camati2016,Goold2014b}.

Irreversibility is traditionally characterized by the concept of entropy production. 
But entropy production is not a physical observable and must therefore be related to observables by means of a theoretical framework \cite{Onsager1931,*Onsager1931a,Machlup1953,DeGroot1961,Tisza1957,Germany1976,Tome2012,Tome2010,Spinney2012,Landi2013b,Spohn1978,Breuer2003,Breuer2007,Deffner2011,Oliveira2016a}. 
For instance,  entropy production can be associated with irreversible work in a work protocol\cite{Crooks1998,*Crooks2000,Jarzynski1997,*Jarzynski1997a} or with the heat exchanged between two systems \cite{Jarzynski2004a}.
A more sophisticated approach is based on the idea of stochastic trajectories \cite{Evans1993,*Evans1994,Gallavotti1995b,*Gallavotti1995}, in which case the entropy production is associated with the ratio between forward and time-reversed path probabilities \cite{Crooks1998,*Crooks2000,Spinney2012,Manzano2017a}.

One aspect that has so far not been explored in detail concerns a understanding of entropy production from the perspective of the environment and the global unitary dynamics of system plus environment. 
This problem was studied in Ref.~\cite{Esposito2010a}, where the authors identified a relation between entropy production and system-environment correlations. 
However, it is generally not fully known what is  the importance of this contribution and what other possible contributions there may exist. 
The problem is also aggravated by the fact that such an answer, by construction, cannot be unique, for two main reasons. 
First, there is an infinite number of environments and interactions which leads to the same reduced dynamics of the system. 
And second, most of these reduced dynamics are obtained by means of a series of approximations (e.g. Born-Markov \cite{Breuer2007}),
which  make it impossible to keep track of the true physical contributions to the entropy production.

The goal of this paper is to address the emergence of entropy production from the perspective of the environment and the system-environment interaction. 
To accomplish this, we consider the simple yet non-trivial model  of a bosonic mode undergoing amplitude damping in contact with a zero-temperature bath, as described by a Lindblad master equation.
The key of our approach is to consider all possible Gaussian dilations \cite{Nielsen} of such dynamics; that is, all possible Gaussian unitary transformations, acting in the Hilbert space of both the system and the environment, which generate exactly the master equation that we aim at addressing. 
This eliminates the arbitrariness concerning the choice of the bath and allows us to avoid the use of any approximations that may hamper our ability to describe the entropy production. 

The Gaussianity of the global dynamics  allows us to construct a theory of entropy production using the idea of Wigner entropy, introduced recently in Ref.~\cite{Santos2017b}. 
This theory has the advantage of operating in quantum phase space, where one may identify quasi-probability currents that represent the microscopic manifestations of the irreversible motion \cite{Spinney2012,Seifert2012,Landi2013b}.
Looking then at  the global unitary dynamics, we are  able to identify two contributions to the entropy production, one related to the creation of mutual information between the system and its bath and the other related to the displacement of the bath from equilibrium. 
Moreover, we  identify what are the irreversible  currents acting within the system \emph{and} the bath, and which are responsible for the emergence of irreversibility on the dynamics of the bath. 
To our knowledge, we are unaware of any papers identifying the role played by these irreversible bath currents in the emergence of irreversibility. 
Finally, we also exploit the fact that our framework is  readily applicable to non-Markovian systems, which allows us to identify such  contributions to the entropy production as witnesses of non-Markovianity from the perspective of the bath.
Potential applications to quantum heat engines are also discussed. 

The choice of studying the zero temperature amplitude damping channel is also motivated by a more fundamental reason. 
Despite being one of the simplest examples of an irreversible process, this problem cannot be described by the standard formalism of entropy production, which uses the von Neumann entropy \cite{Spohn1978,Breuer2003,Breuer2007,Deffner2011}. 
The reason is ultimately related to the divergence of the quantum relative entropy when the reference state becomes pure \cite{Frank2013,Muller-Lennert2013,Audenaert2013,Abe2003} and leads to a divergence of the entropy production  in the limit $T\to 0 $. 
But whether this divergence has a physical significance or not has so far been an open question. 
The results presented in this paper indicate that the divergence of the entropy production at zero temperature is nothing but a  mathematical limitation of the quantum relative entropy.

\emph{The model - }
We consider a bosonic system ($S$) 
with Hamiltonian $H_S = \omega a^\dagger a$, where $a$ ($a^\dag$) is the system annihilation (creation) operator. 
We assume that $S$ is subjected to a zero-temperature amplitude-damping channel  described, in the interaction picture with respect to $H_S$, by the Lindblad master equation
\begin{equation}\label{M}
\frac{\ud \rho_S}{\ud t} = 2\kappa \bigg[ a \rho_S a^\dagger - \frac{1}{2}\{ a^\dagger a, \rho_S\} \bigg],
\end{equation}
where $\kappa$ is the decay rate. 
We work in phase space by introducing the Wigner function $W_S(\alpha,\alpha^*)$ and transforming Eq.~(\ref{M}) into the quantum Fokker-Planck equation
\begin{equation}\label{QFP_S}
\partial_t W_S = \partial_{\alpha} J_S + \partial_{\alpha^*} J_S^*,
\end{equation}
where 
\begin{equation}\label{JS}
J_S(W_S) = \kappa \bigg( \alpha + \frac{\partial_{\alpha^*}}{2}\bigg) W_S.
\end{equation}
Eq.~(\ref{QFP_S}) has the form of a continuity equation, thus allowing us to attribute to $J_S$ the meaning of a current in phase space. 
This is further corroborated by the fact that $J_S$ itself vanishes in the equilibrium state, which in this case is  the vacuum  $W_{S}^\infty = e^{-2|\alpha|^2}/\pi$.

The standard formalism of entropy production, which uses the von Neumann entropy, gives diverging results for this model.
To circumvent this difficulty, we have shown in Ref.~\cite{Santos2017b} that for Gaussian states one could use instead the Wigner entropy $S(W_S) = -\int\ud^2\alpha \; W_S \ln W_S$ (which also coincides with the R\'enyi-2 entropy \cite{Adesso2012}). 
The entropy production associated with Eq.~(\ref{QFP_S}) was then found to be \cite{Santos2017b}:
\begin{equation}\label{Pi_S}
\Pi = -\frac{\ud }{\ud t} S(W_S|| W_{S}^\infty) = \frac{4}{\kappa} \int\ud^2\alpha \frac{|J_S(W_S)|^2}{W_S},
\end{equation}
where $S(W_1||W_2) = \int\ud^2 \alpha W_1 \ln (W_1/W_2)$ is the Wigner relative entropy. 
The second equality in Eq.~(\ref{Pi_S})  establishes a direct relation between irreversibility and the existence of the  current $J_S$. 
In fact, within the classical context, the quantity $J_S/W_S$ is usually interpreted as a velocity in phase space \cite{Tome2010,VandenBroeck2010,Seifert2012}.

\emph{Gaussian dilations - }
We now wish to describe the physics behind Eq.~(\ref{Pi_S}) from the perspective of the global dynamics of the system ($S$) plus environment ($E$). 
To do so, we ask what are the possible dilations which reproduce the full dynamics of Eq.~(\ref{M}) exactly. 
We assume that the environment is bosonic, consisting of a set of modes $b_k$  initially prepared in the global vacuum $|\bm{0}\rangle_E$. 
Moreover, since Eq.~(\ref{M}) is Gaussian preserving, the same must also be  true for the global unitary.
Then, the most general Gaussian Hamiltonian must have the form  \cite{DePaula2014,Rivas2010b, Caruso2008}
\begin{equation}\label{HT}
H_T = \omega a^\dagger a + \sum\limits_{k} \Omega_k b_k^\dagger b_k + \sum\limits_k \gamma_k (a^\dagger b_k + b_k^\dagger a),
\end{equation}
where $\Omega_k$ is the  frequency of mode $k$ and $\gamma_k$ the corresponding coupling constant.
Squeezing terms ($a^\dagger b_k^\dagger$) are not allowed due to the fact that the global vacuum $|0\rangle_S \otimes |\bm{0}\rangle_E$ must be a fixed point of the unitary. 
Moreover, any other quadratic Hamiltonian (for instance containing interactions between the bath modes) can be cast into the form~(\ref{HT}) by a suitable normal mode transformation and renormalization of parameters.

The choice of a Gaussian model is primarily due to its simplicity and tractability. 
Non-Gaussian bosonic models may be treated using the Husimi-Q function with only formal modifications to the approach highlighted above.
Further generalizations can also take place by replacing the dilation approach. 
For instance, the  family of dilations which  reproduce finite temperature  Davies maps  is  the family of thermal operations \cite{Oppenheim2002,*Horodecki2013,*Lostaglio2015}.

The  dynamics generated by Eq.~(\ref{HT}) depends only on two auxiliary functions, $g(t)$ and $f_k(t)$, which satisfy (see supplemental material \cite{SupMat}):
\begin{IEEEeqnarray}{rCl}
\label{g}
\frac{\ud g}{\ud t} &=& - i \sum\limits_k \gamma_k e^{i (\omega-\Omega_k)t} f_k(t),
\\[0.2cm]
\label{f}
\frac{\ud f_k}{\ud t} &=& - i \gamma_k e^{-i(\omega-\Omega_k)t} g(t),
\end{IEEEeqnarray}
with initial conditions $g(0) = 1$ and $f_k(0) = 0$. 
We also assume that $S$ starts in a Gaussian state, which is therefore characterized by the numbers $(\mu, N, M)$, where $\mu = \langle a \rangle_0$, $N = \langle \delta a^\dagger \delta a \rangle_0$ and $M = \langle \delta a \delta a \rangle_0$ (here $\delta a = a - \langle a \rangle$).
The first moments are then  $\langle a \rangle_t = \mu g(t)$ and $\langle b_k \rangle_t = \mu f_k(t)$. 
The expression for the covariance matrix, including all $SE$ correlations, is presented in Ref.~\cite{SupMat}.

Substituting the formal solution of Eq.~(\ref{f}) into~(\ref{g})  yields an integro-differential equation for $g(t)$, which is in general non-Markovian.
All results in this paper will be given in terms of $g(t)$ and therefore hold also in the non-Markovian case. 
The Markovian limit of Eq.~(\ref{M}) corresponds to $g(t) = e^{-\kappa t}$ and is recovered asymptotically via a Wigner-Weisskopf approximation \cite{Weisskopf1930,Scully1997}, after certain assumptions on the spectral density of the system \cite{SupMat}.
We shall refer to this as the time-independent Markovian (TIM) limit.

\emph{Quantum Fokker-Planck equations -}
In the interaction picture, the global Wigner function $W_{SE}(\alpha, \alpha^*, \beta_1, \beta_1^*, \ldots)$  will  satisfy the unitary equation 
\begin{equation}\label{QFP_SE}
\partial_t W_{SE} = \partial_\alpha \mathcal{J}_S + \partial_{\alpha^*} \mathcal{J}_S^* + \sum\limits_k (\partial_{\beta_k}\mathcal{J}_k + \partial_{\beta_k^*}\mathcal{J}_k^* ),
\end{equation}
where $\mathcal{J}_S$ and $\mathcal{J}_k$ are  unitary currents. They can be expressed in terms of the auxiliary functions~(\ref{g}) and (\ref{f}) as
\begin{IEEEeqnarray}{rCl}
\label{JS_SE}
\mathcal{J}_S(W_{SE}) &=& 
 \frac{1}{g^*} (\sum\limits_k \dot{f}_k^* \beta_k) W_{SE},
\\[0.2cm]
\label{Jk_SE}
\mathcal{J}_k(W_{SE}) &=& 
- \frac{\dot{f}_k}{\dot{g}} \mathcal{J}_E(W_{SE}),
\end{IEEEeqnarray} 
where $\dot{f}_k = \ud f_k/\ud t$.  
Moreover, we have defined the global bath currents $\mathcal{J}_E(W_{SE}) = \frac{\dot{g}}{g} \alpha W_{SE}$, which act collectively on all bath modes.

Integrating Eq.~(\ref{QFP_SE}) over the bath degrees of freedom yields a quantum Fokker-Planck equation for $S$, which has the form of Eq.~(\ref{QFP_S}) with the marginal current  $J_S(W_S) = \int \ud^2\bm{\beta} \mathcal{J}_S(W_{SE})$.
As shown in~\cite{SupMat}, this current can be written for an arbitrary number of modes in a time-local form, in the spirt of Ref.~\cite{Hall2014}, as 
\begin{equation}\label{JS_non}
J_S(W_S) = \Gamma(t)  \bigg( \alpha + \frac{\partial_{\alpha^*}}{2}\bigg) W_S,
\end{equation}
where $\Gamma(t) = - \text{Re}(\dot{g}/g)$ \footnote{
In general $g$ will be complex and the imaginary part of $\dot{g}/g$ will give rise to a  Lamb-shift  in the oscillator frequency. 
However, this imaginary term enters into the unitary evolution and therefore does not contribute to any entropic quantities. 
We shall therefore omit this contribution for  the purpose of clarity. However, note that all results in this paper are also valid in the presence of the Lamb-shift.}.
This current has the same form as Eq.~(\ref{JS}) so that $\Gamma(t)$ may be associated with the loss rate in Eq.~(\ref{M}). 
Indeed, in the TIM limit we get precisely $\Gamma(t) = \kappa$. 
Note also that, since our master equation has only a single dissipator, it then follows that one may directly associate Markovianity with the positivity of $\Gamma(t)$ \cite{Hall2014}.

We may also  take the opposite route and trace Eq.~(\ref{QFP_SE}) over the system to obtain a non-Markovian equation for the environment, which has the form
\begin{equation}\label{QFP_E}
\partial_t W_E = \sum\limits_k \partial_{\beta_k} J_k + \partial_{\beta_k^*} J_k^*,
\end{equation}
where $J_k = - (\dot{f}_k/\dot{g}) J_E$ and  
$J_E(W_E) = \int \ud^2 \alpha \mathcal{J}_E(W_{SE})$ is the marginal current, 
which can be written as \cite{SupMat}:
\begin{equation}\label{JE_non}
J_E(W_E) = \dot{g} \bigg\{ \mu - \sum\limits_q (N f_q^* \partial_{\beta_q^*} + M f_q \partial_{\beta_q})\bigg\} W_E.
\end{equation}
The system therefore acts as a  non-Markovian environment for $E$, which introduces displacements, thermal fluctuations and squeezing, depending on the initial conditions $(\mu,N,M)$. 

\emph{Entropy production -}
Having the full solution for $W_{SE}$, we may now proceed to analyze the entropy production from the perspective of the bath. 
Eq.~(\ref{HT}) conserves the total number of quanta in the system. 
This allows one to derive the following entropic conservation law
\begin{equation}\label{conservation}
\frac{\ud S(W_{SE} || W_S^\infty W_E(0))}{\ud t} = 0,
\end{equation}
which means that the entropic distance to the global vacuum remains the same at all times during the evolution.
Using this result one may then express the entropy production rate  $\Pi$ in Eq.~(\ref{Pi_S}) as 
\begin{equation}\label{Pi_E}
\Pi = \frac{\ud \mathcal{I}_{SE}}{\ud t} + \frac{\ud S(W_E||W_E(0))}{\ud t},
\end{equation}
where $\mathcal{I}_{SE} = S(W_{SE}||W_S W_E)$ is the Wigner mutual information between $S$ and $E$~\cite{SupMat}. 
This  shows that the entropy production rate, which is usually expressed as a local quantity of $S$, has two clear contribution: one is a local quantity representing the production of entropy within $E$ and the other is a non-local term related to the rate at which $S$-$E$ correlations build up.
We note that these two mechanisms were also studied 
in Ref.~\cite{Mazzola2012}, where they were related with  the possibility of observing non-Markovianity. 
Eq.~(\ref{Pi_E}) thus holds the potential for  pinpointing the effects of non-Markovianity  in irreversible non-equilibrium processes, a topic of large interest both fundamentally and technologically. 

As a further remark, a similar argument was found in Ref.~\cite{Esposito2010a}, where the entropy production resulting from a non-equilibrium process was ascribed to the difference between the (in general quantum correlated) system-environment state and the tensor product between the reduced state of the system and the equilibrium state of the environment. Eq.~\eqref{Pi_E} clearly identifies both the above contributions to the entropy production, but expresses them from the perspective of the environment, thus providing an original (and indeed fruitful) take to the effects of system-environment interaction. 


We may now express the quantities in Eq.~(\ref{Pi_E}) in terms of the irreversible currents $J_S$ and $J_E$ generated within the system and the environment. 
First, Eq.~(\ref{Pi_S}) is  simply replaced by 
\begin{equation}\label{Pi_SS}
\Pi = -\frac{\ud }{\ud t} S(W_S|| W_{S}^\infty) = \frac{4}{\Gamma} \int\ud^2\alpha \frac{|J_S(W_S)|^2}{W_S},
\end{equation}
which holds for arbitrary time-dependent $\Gamma$.
Next we do the same for the last term in Eq.~(\ref{Pi_E}), which surprisingly  can be cast almost in exactly the same form, as 
\begin{equation}\label{Pi_EE}
\frac{\ud S(W_E||W_E(0))}{\ud t} = \frac{4}{\Gamma} \int \ud^2\bm{\beta} \; \frac{|J_E(W_E)|^2}{W_E}.
\end{equation}
Despite having almost the same structure as Eq.~(\ref{Pi_S}), this result  refers to the contribution of entropy production generated within the bath. 
We therefore see that part of the entropy production is due to the creation of irreversible currents $J_E$ within the environment. 
The remaining part, related to the mutual information, is non-local but may be written using Eqs.~(\ref{Pi_S}), (\ref{Pi_E}) and (\ref{Pi_EE}), as 
\begin{equation}\label{Mutual}
\frac{\ud \mathcal{I}_{SE}}{\ud t} = \frac{4}{\Gamma} \int \ud^2\alpha \ud^2\bm{\beta} \;W_{SE} \bigg\{\frac{|J_S|^2}{W_S^2} - \frac{|J_E|^2}{W_E^2}\bigg\}.
\end{equation}
The creation of mutual information is thus related to the global average mismatch between the phase space velocities in the system ($J_S/W_S$)  and  the bath ($J_E/W_E$).

\emph{Examples -}
The expressions of all entropic quantities appearing in this paper can be written in terms of $\mu$, $N$, $M$ and $g$, and are presented in the Supplemental Material~\cite{SupMat}. 
Here, let us analyze some specific examples. 
A very special case, which is of particular interest, is when the system starts in a coherent state $\rho_S(0) = |\mu\rangle\langle \mu |$ \cite{DePaula2014} (that is, $N = M = 0$).
This situation is  atypical because the solution of  Eq.~(\ref{M}) turns out to also be a coherent state with $\mu_t = \mu g(t)$, which means that the global system  remain in a product state throughout.
Consequently,  $S(W_S)$ and $S(W_E)$ are both constant in time and hence  $\mathcal{I}_{SE} = 0$ throughout the motion. 
Despite being a very particular case, this  example  serves to show that irreversibility may emerge even in the complete absence of  correlations. 
Moreover, it serves as a counterexample to show that the divergence of the standard von Neumann entropy production \cite{Spohn1978,Breuer2003,Breuer2007,Deffner2011}, discussed previously, is not due to any sophisticated feature of the system-bath interaction, but is merely a mathematical limitation of the standard formalism.

\begin{figure}[!b]
\centering
\includegraphics[width=0.22\textwidth]{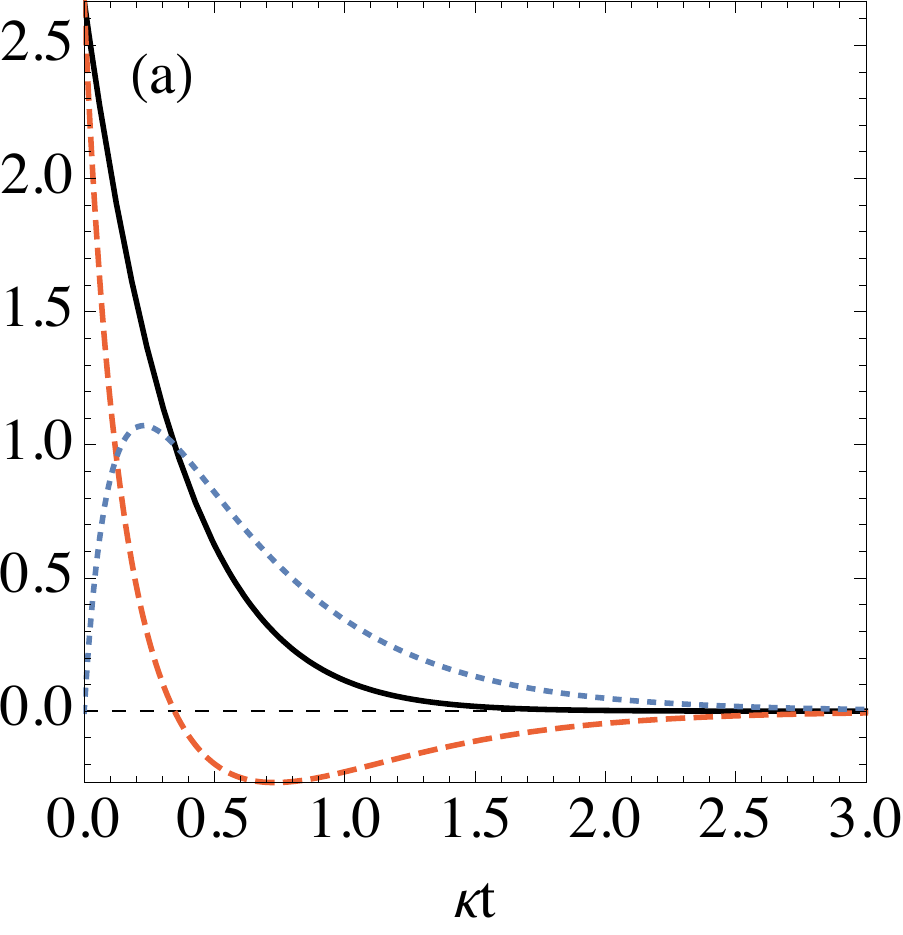}\quad
\includegraphics[width=0.22\textwidth]{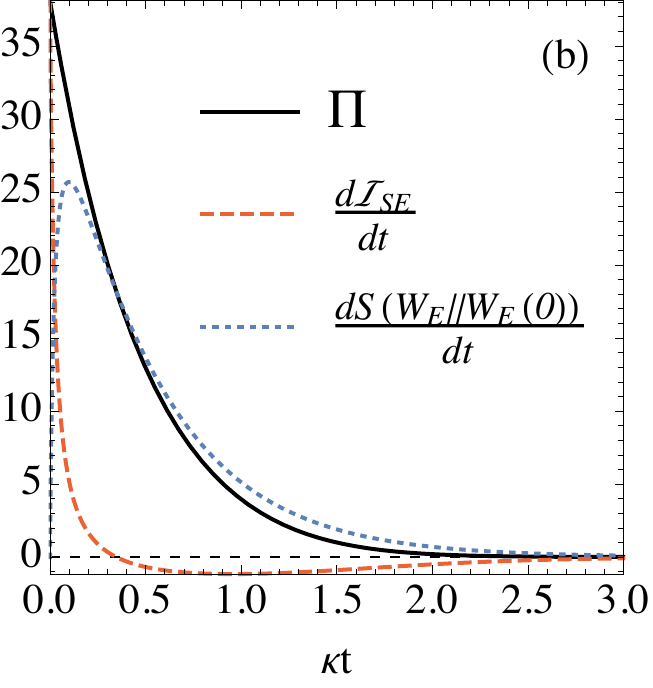}
\caption{\label{fig:thermal}Example of the 3 terms in Eq.~(\ref{Pi_E}) for a thermal state ($\mu = M = 0$) with (a) $N = 1$ and (b) $N = 10$.
The inset in (a) shows the corresponding integrated quantities, from $0$ to $t$. 
}
\end{figure}

Next we consider a thermal initial state where $\mu = M = 0$ and $N = (e^{\beta \omega}-1)^{-1}$ is the mean excitation number for $S$. 
In this case, the state continues to be thermal but with an occupation number $N |g|^2$.
The three quantities appearing in Eq.~(\ref{Pi_E}) are shown in Fig.~\ref{fig:thermal} for $N = 1$ and $N=10$ for the TIM limit.
As can be seen, for larger values of $N$ the contribution to $\ud \mathcal{I}_{SE}/\ud t$ becomes smaller, except at very short times. 
Thus, in the high-temperature limit the correlation between system and bath contributes negligibly to the irreversible behavior.

\emph{Non-Markovianity -}
Our main focus so far has been on the dilations of the Markovian dynamics described in Eq.~(\ref{M}). 
However, all results presented here also hold in the non-Markovian case and therefore provide us with an ideal platform to understand the interplay between the emergence of irreversibility and non-Markovianity.
The evolution of the entropic quantities considered here provide us with several witnesses of non-Markovianity. 
Surprisingly, the mutual information $\mathcal{I}_{SE}$ is not one of them since $\ud \mathcal{I}_{SE}/\ud t$ does not have a well defined sign, even in the Markovian case (see Fig.~\ref{fig:thermal}). 
Instead, one can witness Markovianity by monitoring the distance of $S$ or $E$ from their  respective vacua; that is, $S(W_S||W_S^\infty)$ and $S(W_E||W_E(0))$. 
As an overall Markovian dynamics occurs when $\Gamma(t)>0$, we see from  Eqs.~(\ref{Pi_S}) and (\ref{Pi_EE}) that, in the Markovian case, the system will relax monotonically towards the vacuum, whereas the bath will distance itself from it monotonically.
Thus, any reversal in the velocity at which these processes occur can be take as witnesses  of non-Markovianity. 

We can also link the backflow of information, a key figure of merit of non-Markovianity, with the entropy flux, defined as the mismatch between the entropy production and the total change in the system entropy, $\Pi - \ud S(W_S)/\ud t$.
Using the definition of $\mathcal{I}_{SE}$ in Eq.~(\ref{Pi_E}) one may write $\Phi$ as
\begin{IEEEeqnarray}{rCl}
\Phi &=& \frac{\ud S(W_E)}{\ud t} + \frac{\ud S(W_E||W_E(0))}{\ud t}
= -4 \Gamma \langle a^\dagger a \rangle_t.
\end{IEEEeqnarray}
For finite temperature environments, the entropy flux can be either positive or negative, depending on whether the system was initially warmer or colder than the bath. 
But for a zero-temperature bath, in the Markovian limit the entropy flux can only be from $S$ to $E$. 
A backflow of entropy from $E$ to $S$ can therefore be directly related to a backflow of information.

Finally, we may also witness non-Markovianity  by monitoring the entanglement of the system $S$ with an ancilla $A$~\cite{Rivas2010a}. 
To do that, we consider again the specific example of a thermal state with occupation $N$, but suppose instead that this thermal state actually stems from the two-mode squeezing between the system mode $a$ and an ancila mode $c$. That is, $\rho_{AS}(0) = V |0\rangle_{AS} \langle 0| V^\dagger$, where $V = e^{z(a^\dagger c^\dagger - a c)}$ and $N = \sinh^2(z)$. 
Then, as shown in~\cite{SupMat}, the system-ancilla mutual information may  be related to the other entropic quantities appearing in Eq.~(\ref{Pi_E}), as
\begin{IEEEeqnarray*}{rCl}
\frac{\ud \mathcal{I}_{AS}}{\ud t} &=& - \frac{2N}{N+1} \bigg\{|g|^2 [N(1-|g|^2)+\nicefrac{1}{2}]  \bigg\}\Pi 
\\[0.2cm]
&=& 
- \frac{2N}{N+1} \bigg\{ (1-|g|^2) [N|g|^2+\nicefrac{1}{2}]\bigg\} \frac{\ud S(W_E||W_E(0))}{\ud t}\nonumber
\\[0.2cm]
&=& 
\frac{N}{N+1} (1-2 |g|^2) \frac{\ud \mathcal{I}_{SE}}{\ud t}.
\end{IEEEeqnarray*}
Thus, we see that 
 $\ud \mathcal{I}_{AS}/\ud t$ is related to both $\Pi$ and $\ud S(W_E||W_E(0))/\ud t$ by non-positive pre-factors, so that in the Markovian case $\mathcal{I}_{AS}$ will decay monotonically. 
Conversely,  $\ud \mathcal{I}_{AS}/\ud t$ is linked to  $\ud \mathcal{I}_{SE}/\ud t$ by a factor which does not have a definite sign, therefore showing that the system environment correlations cannot be used as a witness of non-Markovianity.

\emph{Applications - }
Although  we have focused on the conceptual implications of the emergence of irreversibility,  our results also have potential applications, for instance in designing strategies that minimize  losses in quantum heat engines. 
To that end,  suppose that the relaxation process described here actually corresponds to one of the strokes of a heat engine.
The total entropy produced, which is the time integral of Eq.~(\ref{Pi_S}), will depend only on the initial and final states of the system. 
However, from Eq.~(\ref{Pi_E}), we see that this entropy production will partially go to the production of entropy within the environment and partially to the build-up of $SE$ correlations. 
From the perspective of a real machine, therefore, it is desired to minimize the entropy production in the bath, at the cost of increasing the system-bath correlations.
This may be particularly important for finite-sized environments operating continuously, whose  degradation  will be closely related to non-Markovian effects  and could inspire novel types of quantum control methods applicable to the physics of quantum heat engines \cite{DelCampo2014,Abah2016,*Abah2017}.

\emph{Conclusions -} We have made use of a dilation-like approach to study the open  dynamics of a quantum system and  characterize the occurrence of entropy production resulting from a non-equilibrium process. This approach has allowed us to identify two fundamental mechanisms that are responsible for the production of entropy: on one hand, the dynamical bath introduced through the dilation mechanism may get correlated to the system. On the other hand, its state might differ from the equilibrium one in light of its interaction with the system. 
Both features are independently responsible for the emergence of irreversible entropy and can be associated with the existence of probability currents within the system and the environment,  thus playing a key role in the phenomenology of irreversibility. 
It is also very thought-provoking that both mechanisms can be linked to the occurrence of non-Markovian dynamics~\cite{Mazzola2012}, which remarks the relevance of our approach to the characterization of the features of open quantum system, and paves the way to the investigation on the emergence of objective reality through the concept of quantum Darwinism~\cite{Zurek2003b}. The formalization of such a link will be the focus of forthcoming investigations. 

\emph{Acknowledgments -}
GTL would like to acknowledge the S\~ao Paulo Research Foundation, under grant number 2016/08721-7. 
J. P. Santos would like to acknowledge the financial support from the CAPES (PNPD program) for the postdoctoral grant. MP thanks the SFI-DfE Investigator programme, the Royal Society Newton Mobility scheme and the H2020 Collaborative Project TEQ. GTL and MP are supported by FAPESP and Queen's University Belfast through the SPRINT prrogram. 
RD would like to thank CNPq and INCT-IQ for financial support.  

\bibliography{/Users/gtlandi/Documents/library}

\pagebreak
\widetext
\setcounter{equation}{0}
\setcounter{figure}{0}
\setcounter{table}{0}
\setcounter{page}{1}
\makeatletter
\renewcommand{\theequation}{S\arabic{equation}}
\renewcommand{\thefigure}{S\arabic{figure}}
\renewcommand{\bibnumfmt}[1]{[S#1]}
\renewcommand{\citenumfont}[1]{S#1}

\begin{center}
{\bf {\Large Irreversibility at zero temperature from the perspective of the environemnt}}
\vskip0.5cm
{\Large Supplemental Material}
\end{center}


%
%
%
%
\section{Properties of Gaussian Wigner functions}

In this section we provide more explicit formulas for the Gaussian structure of the Wigner functions used throughout the main text. 
Consider a system described by $N$ bosonic modes $a_i$ and let $\bm{R} = (a_1,a_1^\dagger, \ldots, a_N, a_N^\dagger)$. 
We define the mean vector  as 
\begin{equation}\label{mean_vector}
\bm{\mu} = (\langle a_1 \rangle, \langle a_1^\dagger \rangle, \ldots, \langle a_N \rangle, \langle a_N^\dagger \rangle).
\end{equation}
Moreover, the covariance matrix is defined as 
\begin{equation}\label{Theta_def}
\Theta_{i,j} = \frac{1}{2} \langle \{ \delta R_i, \delta R_j^\dagger \} \rangle,
\end{equation}
where $\delta R_i = R_i - \langle R_i \rangle$.
For instance, if $N= 2$, the covariance matrix will have the form 
\begin{equation}
\Theta = \begin{pmatrix}
\langle \delta a_1^\dagger \delta a_1 \rangle + \nicefrac{1}{2} & \langle \delta a_1 \delta a_1 \rangle & \langle \delta a_1 \delta a_2^\dagger \rangle & \langle \delta a_1 \delta a_2 \rangle \\[0.2cm]
\langle \delta a_1^\dagger \delta a_1^\dagger \rangle  & \langle \delta a_1^\dagger \delta a_1 \rangle + \nicefrac{1}{2}  &  \langle \delta a_1^\dagger \delta a_2^\dagger \rangle & \langle \delta a_1^\dagger \delta a_2 \rangle \\[0.2cm]
\langle \delta a_1^\dagger \delta a_2 \rangle & \langle \delta a_1 \delta a_2  \rangle& \langle \delta a_2^\dagger \delta a_2 \rangle + \nicefrac{1}{2} & 
\langle \delta a_2 \delta a_2 \rangle \\[0.2cm]
\langle \delta a_1^\dagger \delta a_2^\dagger \rangle & \langle \delta a_1 \delta a_2^\dagger \rangle & 
\langle \delta a_2^\dagger \delta a_2^\dagger \rangle & \langle \delta a_2^\dagger \delta a_2 \rangle + \nicefrac{1}{2}
\end{pmatrix}.
\end{equation}
The $N$ mode Wigner function is 
\begin{equation}
W(\alpha_1, \ldots, \alpha_N) = \frac{1}{\pi^{2N}} \int \ud^2\lambda_1 \ldots \ud^2 \lambda_N \; e^{-\sum_i (\lambda_i \alpha_i^* - \lambda_i^* \alpha_i)} \tr \bigg\{ \rho e^{\sum\limits_i (\lambda_i a_i^\dagger - \lambda_i a_i)}\bigg\},
\end{equation}
where the integrals are over the entire complex plane of each mode.
For Gaussian states this acquire the specific form 
\begin{equation}\label{W_gaussian}
W = \frac{1}{\pi^N \sqrt{|\Theta|}} \exp\bigg\{ - \frac{1}{2} (\bm{\xi}- \bm{\mu})^\dagger \Theta^{-1} (\bm{\xi} - \bm{\mu})\bigg\}
\end{equation}
where $\bm{\xi} = (\alpha_1, \alpha_1^*, \ldots, \alpha_N, \alpha_N^*)$. 

\subsection{Useful properties of Gaussian Wigner functions}

In this paper we will make extensive use of two expressions that hold for Gaussian Wigner functions. 
They can be derived by simple algebraic manipulations.
First, 
\begin{equation}\label{gaussian_1}
(\xi_i - \mu_i) W = - \sum\limits_{j =1}^{2N} \Theta_{i,j} \partial_{\xi_j^*} W,\qquad i = 1, \ldots, 2N
\end{equation}
Secondly, let $\{1,\ldots, M\}$ be a sub-set of the $N$ modes. 
It then follows that 
\begin{equation}\label{gaussian_2}
\int \ud^2\alpha_1  \ldots \ud^2 \alpha_M \;\; \xi_k W = \begin{cases}
\xi_k \bar{W}, & \text{ if } k \notin \{1, \ldots, 2M\} \\[0.2cm]
\mu_k \bar{W} - \sum\limits_{q=2M+1}^{2N} \Theta_{k,q} \partial_{\xi_q^*} \bar{W},
& \text{ if } k \in \{1,\ldots, 2M\}
\end{cases}
\end{equation}
where 
\begin{equation}
\bar{W}(\alpha_{M+1},\ldots, \alpha_N) = \int \ud^2\alpha_1  \ldots \ud^2 \alpha_M\;\; W
\end{equation}
is the marginalized Wigner function.

\subsection{Wigner entropy and related quantities}

The Wigner entropy is defined as 
\begin{equation}\label{SW}
S(W) = - \int \ud^2 \bm{\alpha} \; W \ln W
\end{equation}
For Gaussian states~(\ref{W_gaussian}) it becomes
\begin{equation}\label{SW_gaussian}
S(W) = N(1 + \ln \pi) + \frac{1}{2} \ln |\Theta|
\end{equation}
Similarly, the Wigner relative entropy between two Wigner functions $W_1$ and $W_2$ is defined as 
\begin{equation}\label{Srel}
S(W_1||W_2) = \int \ud^2\bm{\alpha} \; W_1 \ln W_1/W_2
\end{equation}
For Gaussian states it becomes
\begin{equation}\label{Srel_gaussian}
S(W_1||W_2) = \frac{1}{2} \ln \bigg( \frac{|\Theta_2|}{|\Theta_1|}\bigg) + \frac{1}{2} \tr(\Theta_1 \Theta_2^{-1}) - N + \frac{1}{2} (\bm{\mu}_1 - \bm{\mu}_2)^\dagger \Theta_2^{-1} (\bm{\mu}_1 - \bm{\mu}_2)
\end{equation}
Finally, we define the Wigner mutual information for a bipartite system AB as 
\begin{equation}\label{IAB}
\mathcal{I}_{AB} = S(W_{AB} || W_{A} W_B) = S(W_A) + S(W_B) - S(W_{AB})
\end{equation}
For Gaussian states it becomes, using Eq.~(\ref{SW_gaussian}),
\begin{equation}\label{IAB_gaussian}
\mathcal{I}_{AB} = \frac{1}{2}\ln \bigg( \frac{|\Theta_A| |\Theta_B|}{|\Theta_{AB}|}\bigg)
\end{equation}

%
%
%
%
\section{\label{sec:solution}Full solution of the global dynamics}

We now turn to the specific problem treated in the main text, where a central mode $a$ is allowed to interact with $K$ additional bath modes $b_k$, $k = 1, \ldots, K$. 
The interaction Hamiltonian is 
\begin{equation}\label{H}
H = \sum\limits_k \gamma_k \bigg( a^\dagger b_k e^{i \Delta_k t} + b_k^\dagger a e^{-i \Delta_k t}\bigg),
\end{equation}
where $\Delta_k = \omega - \Omega_k$.

\subsection{Solution in the Heisenberg picture}

In the Heisenberg picture the operators   evolve in time according to 
\begin{IEEEeqnarray}{rCl}
\label{eq_a}
\frac{\ud a(t)}{\ud t} &=& - i \sum\limits_k \gamma_k b_k(t) e^{i \Delta_k t}
\\[0.2cm]
\label{eq_b}
\frac{\ud b_k(t)}{\ud t} &=& - i  \gamma_k a(t) e^{-i \Delta_k t}
\end{IEEEeqnarray}
The formal solution of the latter is 
\begin{equation}\label{b_integro}
b_k(t) = b_k(0) - i \gamma_k \int\limits_0^t  a(t') e^{-i \Delta_k t'} \ud t' .
\end{equation}
Inserting this in Eq.~(\ref{eq_a})  yields the integro-differential equation 
\begin{equation}\label{integro1}
\frac{\ud a(t)}{\ud t} = - i \sum\limits_k \gamma_k b_k(0) e^{i \Delta_k t} - \sum\limits_k \gamma_k^2 \int\limits_0^t a(t') e^{i \Delta_k (t-t')} \ud t'.
\end{equation}
This equation is the basis for treatments of the quantum Langevin equation, with the first term playing the role of the noise operator. 

Instead of proceeding along these lines, we can take a simpler route and notice that, since we assume that the bath  starts in the vacuum, we will only be interested in taking expectation values over  states of the form 
\begin{equation}\label{rho_SE_0}
\rho_{SE}(0) = \rho_S(0) \otimes |\bm{0}\rangle_E \langle \bm{0}|.
\end{equation}
Let us then define the projection operator
\begin{equation}\label{projector}
P = \mathbb{I}_S \otimes |\bm{0}\rangle_E \langle \bm{0}|.
\end{equation}
and the right-projected operators 
\begin{equation}
\tilde{a}(t) = a(t) P, \qquad \tilde{b}_k(t) = b_k(t) P.
\end{equation}
Multiplying Eq.~(\ref{integro1}) by $P$ on the right eliminates the first term, leading to 
\begin{equation}\label{integro2}
\frac{\ud \tilde{a}(t)}{\ud t} = - \sum\limits_k \gamma_k^2\tilde{a}(t')  e^{i \Delta_k (t-t')}.
\end{equation}
Similarly, multiplying Eq.~(\ref{b_integro}) by $P$ on the right yields 
\begin{equation}\label{b_integro2}
\tilde{b}_k(t) = - i \gamma_k \int\limits_0^t  \tilde{a}(t') e^{-i \Delta_k t'} \ud t' .
\end{equation}

Eqs.~(\ref{integro2}) and (\ref{b_integro2}) can now be solved by choosing 
\begin{IEEEeqnarray}{rCl}
\label{a_sol}
\tilde{a}(t) &=& \bigg[ a \otimes  |\bm{0}\rangle_E \langle \bm{0}|\bigg] g(t)
\\[0.2cm]
\label{b_sol}
\tilde{b}_k(t) &=& \bigg[ a \otimes  |\bm{0}\rangle_E \langle \bm{0}|\bigg] f_k(t)
\end{IEEEeqnarray}
where $g(t)$ and $f_k(t)$ are  c-number functions satisfying
\begin{IEEEeqnarray}{rCl}
\label{g_sup}
\frac{\ud g}{\ud t} &=&- \sum\limits_k \gamma_k^2   \; e^{i \Delta_k (t-t')}\; g(t'),
\\[0.2cm]
\label{f_sup}
f_k(t) &=&- i \gamma_k \int\limits_0^t  g(t') e^{-i \Delta_k t'} \ud t' .
\end{IEEEeqnarray}
and subject to the initial conditions $g(0) = 1$ and $f_k(0) = 0$.

We therefore see that the full solution in the Heisenberg picture is completely characterized by the auxiliary functions $g(t)$ and $f_k(t)$. 
For completeness, we also note that instead of~(\ref{g}) and (\ref{f}), we could write 
\begin{IEEEeqnarray}{rCl}
\label{g2}
\frac{\ud g}{\ud t} &=& - i \sum\limits_k \gamma_k f_k(t) e^{i \Delta_k t}
\\[0.2cm]
\label{f2}
\frac{\ud f_k(t)}{\ud t} &=& - i  \gamma_k g(t) e^{-i \Delta_k t}
\end{IEEEeqnarray}
We also mention in passing that $g$ and $f_k$ satisfy
\begin{equation}
|g|^2 + \sum\limits_k |f_k|^2 = 1,
\end{equation}
which is a consequence of the fact that the Hamiltonian~(\ref{H}) conserves the total number of quanta in the system.


We can rewrite Eq.~(\ref{g}) as 
\begin{equation}\label{g2}
\frac{\ud g}{\ud t} = - \int\limits_0^t \mathcal{K}(t-t') g(t') \ud t',
\end{equation}
where $\mathcal{K}$ is the memory Kernel, defined as 
\begin{equation}\label{kernel}
\mathcal{K}(t-t') = \sum\limits_k \gamma_k^2 \; e^{i \Delta_k (t-t')}.
\end{equation}
If one is interested, in particular, in the limit where the bath modes form a continuum, then it is convenient to define the spectral density 
\begin{equation}
J(\Omega) =  2\pi\sum\limits_k \gamma_k^2 \; \delta(\Omega - \Omega_k),
\end{equation} 
which allows us to rewrite Eq.~(\ref{kernel}) as 
\begin{equation}
\mathcal{K}(t-t') = \int\limits_0^\infty J(\Omega) e^{i (\omega - \Omega)(t-t')} \frac{\ud \Omega}{2\pi}
\end{equation}
When $J(\Omega)$ is sufficiently smooth and the central oscillator's frequency $\omega$ is sufficiently large, then  we may employ the Wigner-Weisskopf approximation, which yields
\begin{equation}
\mathcal{K}(t-t')  \simeq 2 \kappa \delta(t-t'),
\end{equation}
where $\kappa = J(\omega)/2$. 
Eq.~(\ref{g2}) then becomes the simple differential equation
\begin{equation}\label{g_mark}
\frac{\ud g}{\ud t} = - \kappa g \qquad \to \qquad g(t) = e^{-\kappa t}
\end{equation}
Thus, in this limit the function $g(t)$ simply relaxes exponentially towards zero. 
We refer to this in the main text as the time-independent Markov (TIM) limit.
Inserting this result in Eq.~(\ref{f}) then yields 
\begin{equation}
f_k(t) = \frac{i\gamma_k}{\kappa + i (\omega - \Omega_k)} \bigg[ e^{-(\kappa + i \Delta_k)t} - 1\bigg].
\end{equation}
As $t\to \infty$ the functions $f_k$ therefore relax to a Lorentzian distribution centered around $\omega$ and of width $\kappa$. 

\subsection{Construction of the covariance matrix}

With the formal solution~(\ref{a_sol}) and (\ref{b_sol}) we may now construct the full mean-vector and covariance matrix of the problem. 
For the initial conditions, we assume that  $\rho_S(0)$ is Gaussian and is thus completely characterized by the three numbers 
\begin{IEEEeqnarray}{rCl}
\mu &=& \langle a \rangle_0,\\[0.2cm]
N &=& \langle \delta a^\dagger \delta a \rangle_0,\\[0.2cm]
M &=& \langle \delta a \delta a \rangle_0.
\end{IEEEeqnarray}
Recall that any Gaussian state of a single mode can be expressed as a displaced squeezed thermal state. 
Hence, it is also convenient to parametrize the parameters $N$ and $M$ as  
\begin{IEEEeqnarray}{rCl}
\label{N}
N+\nicefrac{1}{2} &=& (\bar{n}+\nicefrac{1}{2}) \cosh(2r),\\[0.2cm]
\label{M_Sup}
M &=& (\bar{n}+\nicefrac{1}{2}) e^{i \theta} \sinh(2r).
\end{IEEEeqnarray}
Then $\bar{n}$ represents the thermal occupation of the bath mode, whereas $z = r e^{i \theta}$ is the single-mode squeezing. 

The mean of the annihilation operators will evolve simply as 
\begin{IEEEeqnarray}{rCl}
\langle a \rangle_t &=& \mu  g(t),
\\[0.2cm]
\langle b_k \rangle_t &=& \mu  f_k(t)
\end{IEEEeqnarray}
To compute the covariance matrix, some care must be taken with respect to the ordering of the operators. 
In this sense, the following facts are useful.
First,  the initial state $\rho_{SE}(0)$ in Eq.~(\ref{rho_SE_0}) is invariant under the action of the projector $P$ in Eq.~(\ref{projector}) so $\rho_{SE}(0)  = P \rho_{SE}(0) P$. 
And second,  the operators $a(t)$ and $b_k(t)$ commute (at equal times), but the projected operators $\tilde{a}(t)$ and $\tilde{b}_k(t)$ do not. 
Thus, for instance, we have
\begin{IEEEeqnarray}{rCl}
\langle a b_k^\dagger \rangle &=& \tr\bigg\{ a(t) b_k^\dagger(t) \rho_{SE}(0) \bigg\}
\\[0.2cm]
&=& \tr\bigg\{P b_k^\dagger(t) a(t) P \rho_{SE}(0) \bigg\}
\\[0.2cm]
&=& \tr\bigg\{\tilde{b}_k^\dagger(t) \tilde{a}(t) \rho_{SE}(0) \bigg\}
\\[0.2cm]
&=& \langle a^\dagger a \rangle_0 \;g(t) f_k^*(t)
\end{IEEEeqnarray}

Continuing in this sense, we may now construct the full covariance matrix of S+E. 
We order the operators as $(a,a^\dagger, b_1, b_1^\dagger, \ldots)$. 
Then we parametrize the covariance matrix as 
\begin{equation}\label{sol_second}
\Theta_{SE}(t) = \begin{pmatrix}
\Theta_S 				& 	\Theta_{S,1} 		& \Theta_{S,2} 		& \ldots \\[0.2cm]
\Theta_{S,1}^\dagger 	&	\Theta_{1,1}		& \Theta_{1,2} 		& \ldots \\[0.2cm]
\Theta_{S,2}^\dagger	&	\Theta_{1,2}^\dagger	& \Theta_{2,2} 		& \ldots \\[0.2cm]
\vdots 				&	\vdots			& \vdots			& \ddots
\end{pmatrix}.
\end{equation}
The final solution for each block will then be
\begin{IEEEeqnarray}{rCl}
\label{Theta_S}
\Theta_S(t) &=&
 \begin{pmatrix} 
\langle \delta a^\dagger \delta a \rangle_t + \nicefrac{1}{2} 		&	\langle \delta a \delta a \rangle_t		\\[0.2cm]
\langle \delta a^\dagger \delta a^\dagger \rangle_t				& 	\langle \delta a^\dagger \delta a \rangle_t + \nicefrac{1}{2}  	
\end{pmatrix}
\\[0.2cm]
&=&
 \begin{pmatrix} 
N |g|^2 + \nicefrac{1}{2} 		&	M g^2		\\[0.2cm]
M^* g^{* 2} 				& 	N |g|^2 + \nicefrac{1}{2} 	
\end{pmatrix},
\\[0.2cm]
\Theta_{S,k} &=& 
\begin{pmatrix} 
\langle \delta a \delta b_k^\dagger \rangle_t				&	\langle \delta a \delta b_k \rangle_t		\\[0.2cm]
\langle \delta a^\dagger \delta b_k^\dagger \rangle_t			& 	\langle \delta a^\dagger \delta b_k \rangle_t
\end{pmatrix}
\\[0.2cm]
\label{Theta_Sk}
&=& 
\begin{pmatrix} 
N g f_k^* 					&	M g f_k		\\[0.2cm]
M^* g^* f_k^*				& 	N g^* f_k
\end{pmatrix},
\\[0.2cm]
\Theta_{k,q} &=& 
\begin{pmatrix} 
\langle \delta b_q^\dagger \delta b_k \rangle_t + \delta_{k,q}/2 		&	\langle \delta b_k \delta b_q \rangle_t		\\[0.2cm]
\langle \delta b_k^\dagger \delta b_q^\dagger \rangle_t			& 	\langle \delta b_k^\dagger \delta b_q \rangle_t + \delta_{k,q}/2  	
\end{pmatrix}
\\[0.2cm]
&=&
\begin{pmatrix} 
N f_k f_q^* + \delta_{k,q}/2					&	M f_k f_q		\\[0.2cm]
M^* f_k^* f_q^*				& 	N  f_k^* f_q+ \delta_{k,q}/2
\end{pmatrix},
\label{Theta_E}
\end{IEEEeqnarray}
where, for clarity, we have also written down the corresponding expectation values of each block.

\subsection{Lyapunov equation}

The covariance matrix~(\ref{sol_second})-(\ref{Theta_E}) may also be obtained directly by constructing a Lyapunov equation for $\Theta_{SE}$ using the Hamiltonian~(\ref{H}). 
The equation has the form 
\begin{equation}\label{lyap}
\frac{\ud \Theta_{SE}}{\ud t} = W \Theta_{SE} + \Theta_{SE} W^\dagger,
\end{equation}
where the matrix $W$ is given in block structure as 
\begin{equation}
W = \begin{pmatrix}
0 & \eta_1 & \ldots &\eta_K \\
-\eta_1^\dagger & 0 & \ldots & 0  \\
\vdots & \vdots & & \vdots \\
-\eta_K^\dagger & 0 & \ldots & 0 
\end{pmatrix}
\end{equation}
where
\begin{equation}
\eta_k(t) = -i \gamma_k \text{diag}(e^{i \Delta_k t}, - e^{-i \Delta_k t})
\end{equation}
It is now straightforward to check that Eqs.~(\ref{sol_second})-(\ref{Theta_E}) are indeed the solutions of Eq.~(\ref{lyap}).

%
%
%
%
\section{Evolution of entropic quantities}

Let us now analyze the evolution of some relevant entropic quantities, which can be obtained by using Eqs.~(\ref{sol_second})-(\ref{Theta_E}) in the general formulas~(\ref{SW_gaussian}), (\ref{Srel_gaussian}) and (\ref{IAB_gaussian}). 
The total Wigner entropy of S+E remains constant since the dynamics is unitary. Hence, 
\begin{equation}\label{det_SE}
|\Theta_{SE}(t)| = |\Theta_{SE}(0)| = \frac{(N+\nicefrac{1}{2})^2- |M|^2}{4^K}  = \frac{(\bar{n}+\nicefrac{1}{2})^2}{4^K},
\end{equation}
where, recall, $\bar{n}$ is the thermal occupation number defined in Eq.~(\ref{N}).
When $\bar{n} = 0$ the global S+E state will be pure. 

The determinants of the system covariance matrix $\Theta_S$ and the environment covariance matrix $\Theta_E$ evolve as 
\begin{IEEEeqnarray}{rCl}
\label{det_S}
|\Theta_S(t)| &=& (N |g|^2 + \nicefrac{1}{2})^2 - |M|^2 |g|^4
\\[0.2cm]
\label{det_E}
|\Theta_E(t)| &=& \frac{1}{2^{2K-2}} \bigg\{ 
[N (1-|g|^2) + \nicefrac{1}{2}]^2 - |M|^2 (1-|g|^2)^2\bigg\}
\end{IEEEeqnarray}
Eqs.~(\ref{det_SE})-(\ref{det_E}) can then be used in Eqs.~(\ref{SW_gaussian}) and (\ref{IAB_gaussian}) to completely determine the time evolution of $S(W_S)$, $S(W_E)$ and $\mathcal{I}_{SE}$ [cf. Eq.~(\ref{IAB_gaussian})].
We note that in the Markovian case the mutual information starts from zero, reaches a maximum and then decays back to zero as $t\to \infty$. 


As for relative entropies, the two most important ones are (see main text for a proper context)
\begin{equation}\label{S_WS_WSinf}
S(W_S|| W_S^\infty) = 2(N + |\mu|^2)|g|^2 - \frac{1}{2} \ln \bigg\{(N |g|^2 + \nicefrac{1}{2})^2 - |M|^2 |g|^4 \bigg\} - \ln 2 
\end{equation}
and 
\begin{equation}\label{S_WE_WE0}
S(W_E|| W_E(0)) = 2(N + |\mu|^2)(1-|g|^2) - \frac{1}{2} \ln \bigg\{[N (1-|g|^2) + \nicefrac{1}{2}]^2 - |M|^2 (1-|g|^2)^2 \bigg\} - \ln 2^{K} 
\end{equation}
Differentiating Eq.~(\ref{S_WS_WSinf}) with respect to time then yields an explicit formula for the entropy production rate 
\begin{equation}\label{Pi}
\Pi = - \frac{\ud S(W_S||W_S^\infty)}{\ud t} = 4 |g|^2 \Gamma \bigg\{ |\mu|^2 + N -\frac{1}{4} \frac{N + 2(N^2-|M|^2) |g|^2}{(N|g|^2+\nicefrac{1}{2})^2 - |M|^2 |g|^4}\bigg\}
\end{equation}
where, recall, $\Gamma = - \text{Re}(\dot{g}/g)$.
Recall also that from the main text, we divide $\Pi$ as
\begin{equation}
\Pi = \frac{\ud \mathcal{I}_{SE}}{\ud t} + \frac{\ud S(W_E||W_E(0))}{\ud t} 
\end{equation}
Differentiating Eq.~(\ref{S_WE_WE0}) with respect to time then gives
\begin{equation}\label{PiE}
\frac{\ud S(W_E||W_E(0))}{\ud t}  = 4 |g|^2 \Gamma  \bigg\{|\mu|^2 + N - \frac{N + 2 (N^2-|M|^2)(1-|g|^2)}{[N (1-|g|^2) + \nicefrac{1}{2}]^2 - |M|^2 (1-|g|^2)^2}\bigg\}
\end{equation}
The difference between Eq.~(\ref{Pi}) and (\ref{PiE}) then gives $\ud \mathcal{I}_{SE}/\ud t$.

%
%
%
%
\section{Reduced Fokker-Planck equations}

In this section we present the details on how to obtain reduced Fokker-Planck equations for the system and the environment, represented by Eqs.~(12) and (14) of the main text. 

We start with the global unitary equation for $W_{SE}$:
\begin{equation}\label{QFP_SE_SUp}
\partial_t W_{SE} = \partial_\alpha \mathcal{J}_S + \partial_{\alpha^*} \mathcal{J}_S^* + \sum\limits_k (\partial_{\beta_k}\mathcal{J}_k + \partial_{\beta_k^*}\mathcal{J}_k^* ),
\end{equation}
where
\begin{IEEEeqnarray}{rCl}
\label{JS_SE}
\mathcal{J}_S(W_{SE}) &=& 
 \frac{1}{g^*} (\sum\limits_k \dot{f}_k^* \beta_k) W_{SE},
\\[0.2cm]
\label{Jk_SE}
\mathcal{J}_k(W_{SE}) &=& 
- \frac{\dot{f}_k}{\dot{g}} \mathcal{J}_E(W_{SE})
\end{IEEEeqnarray} 
and
\begin{equation}\label{JE_SE}
\mathcal{J}_E(W_{SE}) = \frac{\dot{g}}{g} \alpha W_{SE},
\end{equation}

\subsection{Reduced current for S}

Integrating Eq.~(\ref{QFP_SE}) over all bath modes $\beta_k$ yields a reduced Fokker-Planck equation for the system's Wigner function $W_S$.
Upon integration, the terms in the RHS which depend on $\partial_{\beta_k}$ will vanish, leaving us with 
\begin{equation}\label{QFP_S_Sup}
\partial_t W_S = \partial_{\alpha} J_S + \partial_{\alpha^*} J_S^*,
\end{equation}
where
\begin{equation}\label{JS_non}
J_S(W_S) = \int \ud^2\bm{\beta} \; \mathcal{J}_S(W_{SE})
\end{equation}
We now use Eq.~(\ref{gaussian_2}) to write 
\begin{IEEEeqnarray*}{rCl}
\int \ud^2\bm{\beta} \; \beta_k W_{SE} &=& \bigg\{\mu f_k - (\Theta_{SE})_{2k+1,1} \partial_{\alpha^*} - (\Theta_{SE})_{2k+1,2} \partial_\alpha \bigg\} W_S
\\[0.2cm]
&=& f_k \bigg\{\mu  -N g^* \partial_{\alpha^*} - M g \partial_\alpha \bigg\} W_S
\end{IEEEeqnarray*}
where in the second line we used Eq.~(\ref{Theta_Sk}). 
Substituting this in Eqs.~(\ref{JS_SE}) and (\ref{JS_non}), and using Eq.~(\ref{g2}), yields
\begin{equation}\label{JS_reduced_1}
J_S = - \dot{g} \bigg\{ \mu - N g^* \partial_{\alpha^*} - M g \partial_\alpha \bigg\} W_S
\end{equation}
Finally, we use Eq.~(\ref{gaussian_1}) for $W_S$ to establish the identity 
\begin{IEEEeqnarray*}{rCl}
\alpha W_S &=& \mu g W_S - (\Theta_S)_{1,1} \partial_{\alpha^*} W_S - (\Theta_S)_{1,2} \partial_\alpha W_S 
\\[0.2cm]
&=& \mu g W_S - (N |g|^2 + \nicefrac{1}{2}) \partial_{\alpha^*} W_S - M g^2 \partial_\alpha W_S
\end{IEEEeqnarray*}
Rearranging this allows us to write 
\begin{equation}
g \bigg\{ \mu - N g^* \partial_{\alpha^*} - M g \partial_\alpha \bigg\} W_S = \bigg(\alpha + \frac{\partial_{\alpha^*}}{2}\bigg) W_S
\end{equation}
which, upon substitution on Eq.~(\ref{JS_reduced_1}), yields 
\begin{equation}\label{JS_reduced_1}
J_S = - \frac{\dot{g}}{g} \bigg(\alpha + \frac{\partial_{\alpha^*}}{2}\bigg) W_S
\end{equation}
which is Eq.~(12) in the main text. 

\subsection{Reduced current for E}

We now repeat the procedure for the environment. 
Integrating Eq.~(\ref{QFP_SE}) with respect to $\alpha$ destroys the first term in the RHS, leaving us with 
\begin{equation}\label{QFP_E}
\partial_t W_E = \sum\limits_k \partial_{\beta_k} J_k + \partial_{\beta_k^*} J_k^*,
\end{equation}
where $J_k = - (\dot{f}_k/\dot{g}) J_E$ and 
\begin{equation}
J_E = \int \ud^2\alpha \; \mathcal{J}_E(W_{SE})
\end{equation}
Based on Eq.~(\ref{JE_SE}) we again use Eq.~(\ref{gaussian_2}), which yields 
\begin{equation}
\int \ud^2\alpha \; \alpha W_{SE} = \mu g W_E - \sum\limits_{k=1}^K \bigg\{ N g f_k^* \partial_{\beta_k^*} + M g f_k \partial_{\beta_k} \bigg\} W_E
\end{equation}
Combining this with Eq.~(\ref{f2}) then gives the desired result, 
\begin{equation}
J_E(W_E) = \dot{g} \bigg\{ \mu - \sum\limits_q (N f_q^* \partial_{\beta_q^*} + M f_q \partial_{\beta_q})\bigg\} W_E.
\end{equation}
which is Eq.~(14) of the main text.

%
%
%
%
\section{Correlation between a system and an ancila}

Now let us consider, in addition to the system mode $a$, an ancila mode $c$. 
Suppose that the initial state of S and A is a two-mode squeezed state of the form 
\[
\rho_{AS}(0) = V |0\rangle_{AS} \langle 0| V^\dagger
\]
where $V = e^{z(a^\dagger c^\dagger - a c)}$ and, without loss of generality, we take $z$ to be real.  
For concreteness, we parametrize $z$ such that  $\sinh^2(z)=N$. 
The covariance matrix associated with the state $\rho_{AS}$ is, with the ordering $(c,c^\dagger, a,a^\dagger)$:
\begin{equation}
\Theta_{AS}(0) = 
\begin{pmatrix}
N+\nicefrac{1}{2} 	&		0		&		0		&		\sqrt{N(N+1)}	\\[0.2cm]
0				&   N+\nicefrac{1}{2} &	\sqrt{N(N+1)}	&		0			\\[0.2cm]
0				&     \sqrt{N(N+1)}     &    N+\nicefrac{1}{2}&    		0			\\[0.2cm]
\sqrt{N(N+1)}		&	0			&		0		& N+\nicefrac{1}{2}
\end{pmatrix}
\end{equation}
Using the results of Sec.~\ref{sec:solution} it is now straightforward to compute the time evolution of the global A+S+E system. 
For our purposes, we are now only interested in the correlations between A and S, so we focus on $\Theta_{AS}(t)$, which is given by 
\begin{equation}
\Theta_{AS}(t) = 
\begin{pmatrix}
N+\nicefrac{1}{2} 	&		0		&		0		&		\sqrt{N(N+1)}g	\\[0.2cm]
0				&   N+\nicefrac{1}{2} &	\sqrt{N(N+1)}g^*	&		0			\\[0.2cm]
0				&     \sqrt{N(N+1)}g     &    N|g|^2+\nicefrac{1}{2}&    		0			\\[0.2cm]
\sqrt{N(N+1)}g^*		&	0			&		0		& N|g|^2+\nicefrac{1}{2}
\end{pmatrix}
\end{equation}
The mutual information between A and S is now readily found using Eq.~(\ref{IAB_gaussian}):
\begin{equation}
\mathcal{I}_{AS} = \ln \bigg\{ 2 \frac{(N+\nicefrac{1}{2})(N |g|^2 + \nicefrac{1}{2})}{N(1-|g|^2) + \nicefrac{1}{2}}\bigg\}
\end{equation}
Differentiating with respect to time also yields
\begin{equation}
\frac{\ud \mathcal{I}_{AS}}{\ud t} = - \Gamma \frac{|g|^2N(N+1)}{(N|g|^2 + \nicefrac{1}{2})(N(1-|g|^2) + \nicefrac{1}{2})}
\end{equation}
which can now be compared with the other entropic quantities of interest. 

\end{document}